\documentclass[prb,twocolumn]{revtex4-1}
\usepackage{amsmath}
\usepackage{amsfonts}
\usepackage{amssymb}  
\usepackage{graphicx}
\usepackage[pdftex,colorlinks=true,linkcolor=blue,citecolor=blue,urlcolor=blue,filecolor=blue,breaklinks=true]{hyperref}  
\usepackage{url}

\newcommand{\cdop}[1]{\mathop{\hat{c}^{\dagger}_{#1}}}
\newcommand{\cop}[1]{\mathop{\hat{c}^{\vphantom \dagger}_{#1}}}

\begin{document}

\title{Visualizing topological transport}

\author{Mariya A.~Lizunova}
\affiliation{Institute for Theoretical Physics, Utrecht University, Princetonplein 5, 3584 CC Utrecht, The Netherlands}
\affiliation{Institute for Theoretical Physics Amsterdam, University of Amsterdam, Science Park 904, 1098 XH Amsterdam, The Netherlands}

\author{Samuel Kuypers}
\author{Bernet Meijer}
\affiliation{Institute for Theoretical Physics Amsterdam, University of Amsterdam, Science Park 904, 1098 XH Amsterdam, The Netherlands}

\author{Ana Silva}
\affiliation{Department of Physics of Complex Systems, Weizmann Institute of Science, Rehovot 76100, Israel}

\author{Jasper van Wezel}
\email{vanwezel@uva.nl}
\affiliation{Institute for Theoretical Physics Amsterdam and Delta Institute for Theoretical Physics, University of Amsterdam, Science Park 904, 1098 XH Amsterdam, The Netherlands}

\date{\today}

\begin{abstract}
We present a mathematically simple procedure for explaining and visualizing the dynamics of quantized transport in topological insulators. The procedure serves to illustrate and clarify the dynamics of topological transport in general, but for the sake of concreteness, it is phrased here in terms of electron transport in a charge-ordered chain, which may be mapped exactly onto transport between edge channels in the Integer Quantum Hall Effect. It has the advantage that it allows a direct visualization of the real-space and real-time evolution of the electronic charges throughout the topological pumping cycle, thus demystifying how charge flows between remote edges separated by an insulating bulk, why the amount of transported charge is given by a topological invariant, and how continuous driving yields a discrete, quantized amount of transported charge.
\end{abstract}

\maketitle

\section{Introduction}
Topology has, over the past decades, taken centre stage alongside symmetry as one of the basic organizing principles of condensed matter physics. As with symmetry, the predictive power associated with topology can be enormous. For example, when electrons confined to move in two dimensions are exposed to a perpendicular magnetic field, they experience the Quantum Hall Effect,\cite{IQHE} in which the transverse conductance becomes equal to precisely $e^2/h$ times an integer. The quantization of the transverse conductance is exact, and independent of how the electron gas is realized experimentally.\cite{TKNN,Avron1985} This is possible, because the integer in the value of the transverse conductivity is a topological quantum number that counts the integer number of conduction channels along the edge of the Quantum Hall material, rather than any microscopic property of the bulk. Predictions of the transverse conductance have been verified to one part in a billion, and in fact the Quantum Hall Effect now serves as a standard for resistance calibration.\cite{IQHEstandard} 

After topology was introduced into condensed matter physics as the theoretical explanation underlying the Integer Quantum Hall Effect (IQHE),\cite{TKNN,Avron1985} many other phases of matter were found to similarly display some type of precisely quantized transport, due to the presence of topological quantum numbers. These include the Fractional Quantum Hall Effect (FQHE),\cite{LaughlinFQHE,TsuiFQHE} Quantum Spin Hall Effect (QSHE),\cite{Haldane1988,QSHE} and more generally Topological Insulators (TIs), semi-metals, and superconductors.\cite{KM,Bernevig,FKM,Armitage2018,TopoSC} Topology has thus become one of the corner stones of modern condensed matter physics.\cite{TIreview2010,TIreview2016} Moreover, topological order has been suggested to open the way towards various applications, including dissipationless topological transport, fault-tolerant quantum computation, and the engineering of spin liquid phases of matter.\cite{TIreview2010,TIreview2016}

Here, we will focus on topological insulators, which include the integer Quantum Hall state. The hallmark of a material being a TI, is that even though it does not conduct electricity through its bulk, there necessarily are robust conducting states along its edges. The topological nature of these edge states is seen most clearly through the phenomenon of topological transport or topological pumping.\cite{QAPT,Thouless,QAPTColdAtoms} Its idea is easily formulated. By periodically changing some driving force, particles are transferred from one edge of the TI to the opposite side. This transport is \emph{quantized}, in the sense that for every period of the pumping cycle, precisely an integer number of particles will move between edges.\cite{QAPT,Thouless} It is \emph{topological}, because the discrete number of relocated particles is independent of the details of both the system and the driving. As long as the driving is smooth enough not to cause a phase transition in the material, the number of transferred particles will always be the same.\cite{Jorrit2017} In fact, the topological nature of the pumping can also be seen as an example of the so-called bulk-boundary correspondence, since the integer topological quantum number describing the number of particles relayed between opposing edges, can be calculated entirely in terms of the electron wave functions in the bulk of the TI.\cite{TKNN,Laughlin} Topological pumps thus bring together all the main players in the modern understanding of topological matter: quantized conductance, topological quantum numbers, edge states, robustness to perturbations, and the bulk-boundary correspondence.\cite{Flicker,ourPRB} Moreover, the topological pump provides a simple and accessible (thought) experiment that can be easily introduced even in the early stages of a physics curriculum.

Unfortunately, the intuitive idea of what topological transport entails, is not easily translated into an equally accessible mathematical description of the pumping process in any explicit model.\cite{ourPRB} Several discussions of how the IQHE and topological transport can be introduced into classrooms do exist,\cite{ajp1,ajp2,ajp3,ajp4,ajp5} but these focus either on particular realizations of the IQHE or TI phases that are not easily related to other forms of topological order;\cite{ajp1,ajp2} or on  phenomenological consequences of quantized edge transport that do not clearly exhibit the connection to the topological bulk;\cite{ajp3,ajp4} or on practical experiments that showcase the effects of topology but that are not easily captured in a mathematical description.\cite{ajp5} Here, we remedy this situation by presenting a particular topological pumping process that emphasizes the roles of the key players (the topological invariant, the pumping process, and the connection between opposing edges) and that allows for a straightforward and direct visualization of the particle transport throughout the pumping cycle. The particular model we consider is that of a one-dimensional chain of atoms with periodic variations in its density of electrons. The analysis of the spectrum, the calculation of a topological quantum number, the simulation and visualization of electronic eigenfunctions, and the identification of edge state dynamics are all mathematically accessible in this model. Moreover, the model can be precisely mapped onto the more standard, but more involved, example of topological transport in the IQHE, and it can be straightforwardly generalized to visualize topological transport in other types of TI as well.

\section{The charge-ordered chain}
\begin{figure}[t]
\begin{center}
\includegraphics[width=\columnwidth]{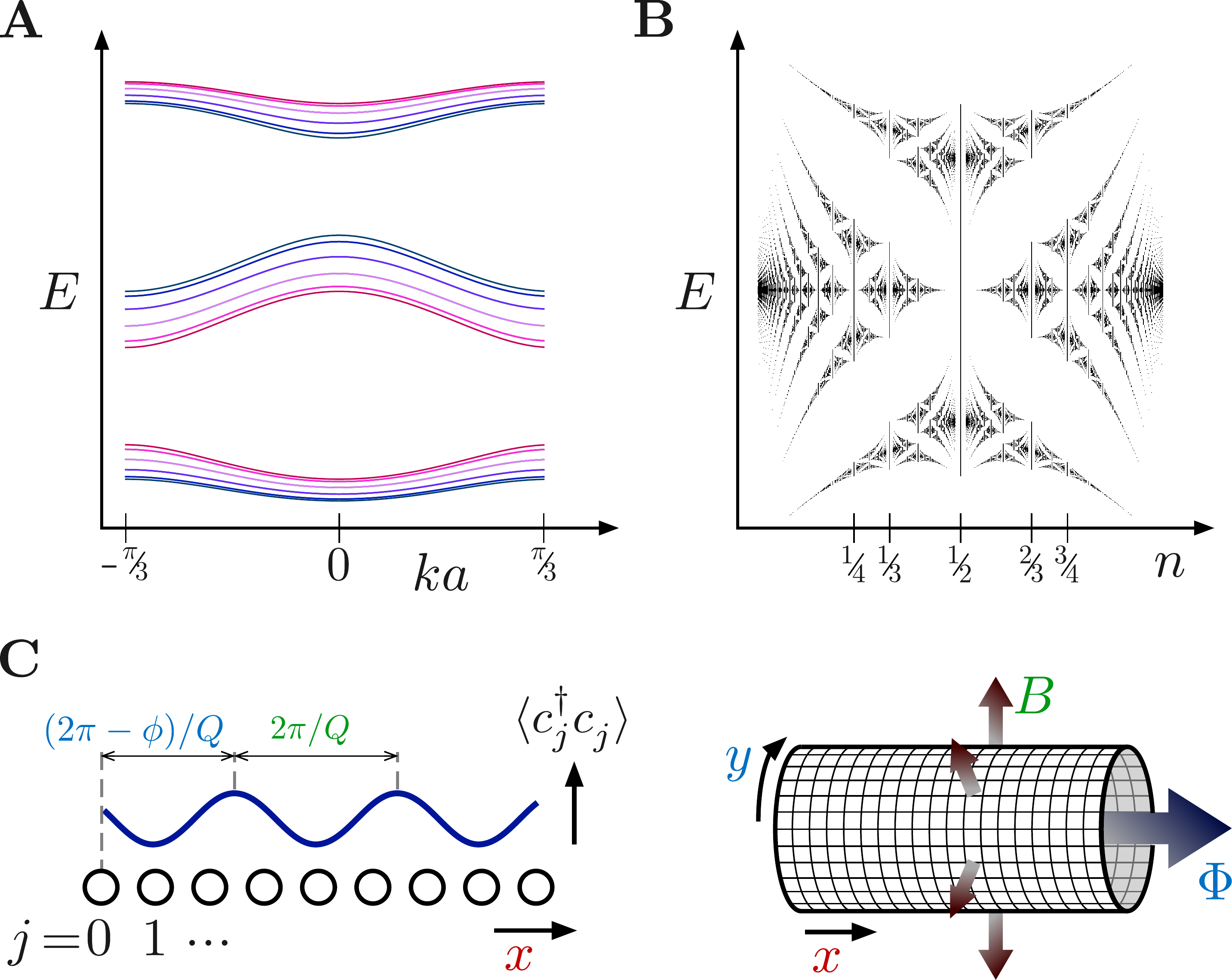}
\end{center}
\caption{\label{fig1}{\bf A} The energy as a function of wave number for the mean-field CDW with periodic boundary conditions. Different colors correspond to different values of $\phi$, ranging between zero and $2\pi$. {\bf B} The possible energies of the mean-field CDW as a function of filling fraction. For each $n$, ranges are indicated for all energies found as $k$ is varied between $-\pi/(3a)$ and $\pi/(3a)$, and $\phi$ is varied between zero and $2\pi$. The resulting figure is known as Hofstadter's butterfly~\cite{Hofstadter}, and was first found in a tight-binding model for the IQHE. {\bf C} Pictorial representation of the correspondence between the mean-field CDW and a tight-binding model for the IQHE.\\
To calculate the energies of the CDW, we considered $50$ $k$-points, and used the (arbitrary) parameter values $V \Delta=0.5 t$, and $Q=2\pi/(3a)$. The phase $\phi$ was varied between zero and $2\pi$ in ten steps. The Hofstadter butterfly contains all fractions $n=p/q$ with $p$ and $q$ co-prime integers and $q$ ten or less. For each $n$, energies were calculated for $200$ steps in both $k$ and $\phi$.}
\end{figure}

As a basic setting for visualizing topological transport, we consider electrons on a one-dimensional chain of atoms, shown in Fig.~\ref{fig1}C. We ignore the spin of the electrons, but do consider a repulsive (Coulomb) interaction between electrons localized on neighboring sites. The Hamiltonian describing this system is:
\begin{align}
  \hat{H} = \sum_{j=0}^{N-1} \left\{ -t \left(\cdop{j} \cop{j+1} +\cdop{j+1} \cop{j} \right)  +V \cdop{j} \cop{j} \cdop{j+1} \cop{j+1} \right\}.
  \label{Hcdw}
\end{align}
Here, the operators $\cdop{j}$ and $\cop{j}$ respectively create and annihilate an electron at position $x = j a$, where $a$ is the lattice constant and $j$ an integer site label. Notice that for now, we will use periodic boundary conditions, so that the labels $j$ and $j+N$ correspond to the same site. The first term in the Hamiltonian, proportional to $t$, describes the tunneling of electrons between neighboring sites. The second term, proportional to $V$, accounts for the nearest-neighbor Coulomb interaction. Both $t$ and $V$ are taken to be positive. Weak impurities could be added to the model of Eq.~\eqref{Hcdw} using a random on-site potential,\cite{ourPRB} but we will not consider this aspect here.

As first proposed by Peierls,\cite{Peierls,Frohlich,Kohn} and observed in many real and artificial materials,\cite{Little,Brazovskii,Kraus1,Gruner} interacting electrons in a one-dimensional chain are expected to spontaneously organize into a spatially modulated pattern at low temperatures. This so-called charge-density wave (CDW) may be described by assuming that the expectation value of electron density, will be of the form:
\begin{align}
  \langle \cdop{j} \cop{j} \rangle = \rho_0 + \rho(j) = \rho_0 + \Delta \cos(Q j a + \phi).
\end{align}
Here $\rho_0$ is the average electron density. If the amplitude $\Delta$ is non-zero, there will be a CDW in the chain. The wave number of the CDW is $Q=n \cdot 2 \pi/a$. It is determined by the (fractional) average number of electrons per site, or filling fraction $n=p/q$, with $p$ and $q$ co-prime integers. The phase $\phi$ determines the position of the CDW with respect to the atoms in the chain, and varying $\phi$ corresponds to sliding the charge modulation along the chain.\cite{Gruner} In practice, such a sliding motion may be induced by an applied electric field, if it is sufficiently strong.\cite{Bardeen,Fukuyama76,LeeRice}

In the Supplementary Material,\cite{SupMat} we describe how assuming the electron density to be described by Eq.~\eqref{Hcdw} leads to a simplified (so-called mean-field) form of the Hamiltonian, which can be written as:
\begin{align}
  \hat{H}_{\text{MF}} = \sum_{j=0}^{N-1} \left\{ -t (\cdop{j} \cop{j+1} +\cdop{j+1} \cop{j} )+ 2V \rho(j) \cdop{j} \cop{j} \right\}.
  \label{HcdwMF}
\end{align}
Assuming periodic boundary conditions, the Hamiltonian is most conveniently written in terms of electrons whose wave functions are plane waves, rather than the electrons with strictly localized wave functions created by $\cdop{j}$. We therefore define the operator $\cdop{k}$, which creates a plane-wave electron with wave number $k$. It can be written in terms of localized states by the relation $\cdop{k}=\sqrt{1/N}\sum_j e^{-ikja} \cdop{j}$. In the Supplementary Material,\cite{SupMat} we use this definition to rewrite the Hamiltonian, and we show that it is equal to:
\begin{align*}
\hat{H}_{\text{MF}} = \sum_{0 \le k < 2 \pi / a} \left\{ \frac{1}{2}\epsilon_{k}^{\phantom \dagger}\hat{c}_{k}^{\dagger}\hat{c}_{k}^{\phantom \dagger}+ V \Delta^{\phantom 
\dagger} e^{i \phi} \hat{c}_{k}^{\dagger}\hat{c}_{k+Q}^{\phantom \dagger}+H.c.\right\}.
\end{align*}
Here, $\epsilon_{k}=-2t \cos\left(k a\right) $ is the energy that a plane-wave electron with wave number $k$ would have in the absence of any Coulomb interaction. To see the effect of the non-zero Coulomb interactions, we can write the Hamiltonian in matrix form:\cite{Flicker}
\begin{align}
\hat{H}_{\text{MF}} &=\sum_{0 \le k < 2\pi / qa} \left(\hat{c}_{k+Q}^{\dagger},\hat{c}_{k+2Q}^{\dagger}  \ldots \hat{c}_{k+qQ}^{\dagger}\right)H_{k}\left(\begin{array}{c} 
\hat{c}_{k+Q}\\ \hat{c}_{k+2Q}\\ \vdots\\ \hat{c}_{k+qQ} \end{array}\right) \notag \\
~&~\notag\\
H_{k} &= \left(\begin{array}{cccccc} \epsilon_{k+Q}^{\phantom *} & \tilde{\Delta}^{\phantom *} & 0 & \ldots & 0 & \tilde{\Delta}^{*} \\
 \tilde{\Delta}^{*} & \epsilon_{k+2Q}^{\phantom *} & \tilde{\Delta}^{\phantom *} & 0 & \ldots & 0 \\
0 & \tilde{\Delta}^{*} & \ddots & \ddots &  & \vdots \\
\vdots & 0 & \ddots &  &  & 0 \\
0 & \vdots &  &  &  & \tilde{\Delta}^{\phantom *} \\
 \tilde{\Delta}^{\phantom *} & 0 & \ldots & 0 & \tilde{\Delta}^{*} & \epsilon_{k+qQ}^{\phantom *} \end{array}\right).
\label{HmfMatrix}  
\end{align}
The parameter $\tilde{\Delta}$ in this expression equals $V\Delta e^{i \phi}$, while $q$ is the denominator in the filling fraction $n=p/q$, and $k+qQ=k$ owing to the periodicity of wave numbers in a discrete atomic chain. Note that $q=2$ is a special case,\cite{Flicker} in which the Hamiltonian cannot be written in the form of Eq.~\eqref{HmfMatrix}, and we will not consider it here.
%

\section{Comparison to the IQHE}
Numerically finding the eigenvalues of the matrix $H_k$ yields the energy of electrons with wave number $k$ for any given value of the CDW phase $\phi$, as shown in Fig.~\ref{fig1}A. Collecting the eigenvalues associated with all possible choices for $k$ and $\phi$ at a given filling fraction $n$, and plotting them as dots in the plane of energy versus filling fraction, yields a version of the famous Hofstadter butterfly spectrum,\cite{Hofstadter,ChangNiu1,Flicker} shown in Fig.~\ref{fig1}B. For the specific case $V \Delta=t$, Eq.~\eqref{HmfMatrix} becomes equivalent to the matrix form of Harper's equation, applied by Hofstadter to model electrons in a two-dimensional plane subjected to a strong perpendicular magnetic field,\cite{Hofstadter} which then display the IQHE. We thus find that the physics of sliding CDW may be mapped onto that of the IQHE.\cite{ChangNiu1,Flicker} To make the correspondence exact, we should consider rolling up the two-dimensional plane of the IQHE into a cylinder, as shown in Fig.~\ref{fig1}C. The magnetic field strength perpendicular to the surface of the IQHE cylinder then correspond to the filling fraction of the CDW chain, and hence to the CDW wave number $Q$.\cite{Flicker,Loss1,Hofstadter} Under the same mapping, the phase $\phi$ of the CDW translates to a magnetic flux threading the quantum Hall cylinder,\cite{Laughlin} while the spatial coordinate of the CDW chain is directly related to the spatial coordinate parallel to the axis of the IQHE cylinder. The mapping is indicated schematically in Fig.~\ref{fig1}C.

In the semi-classical picture of the IQHE, electrons in a two-dimensional plane are forced by the perpendicular magnetic field to move in cyclotron orbits that are much smaller than the spatial extent of the system. Electrons in the bulk of the plane can therefore not conduct electricity.  Charge transport will be possible only when we consider the IQHE on a surface with boundaries. The conductance in that case takes place along the edges of the system, is strictly quantized, topological in nature, and can be calculated from a bulk topological quantum number.\cite{TKNN,Laughlin} This is the main manifestation of the celebrated bulk-boundary correspondence in the IQHE, and in TIs in general.

Like the IQHE, the sliding CDW system is strictly insulating for all values of $\phi$, as long as periodic boundary conditions are applied. When we consider a finite chain with open boundaries, edge states that are localized at the ends of the chain appear. As in the IQHE, the value of the edge state conductance in an open chain may be determined in terms of a topological quantum number, which is a single integer number characterising the energy eigenvalues in the bulk, periodic chain.\cite{Thouless,ChangNiu1,Flicker,ourPRB} We discuss in appendix~\ref{appA} how to calculate the topological quantum number associated with the Hamiltonian in Eq.~\eqref{HmfMatrix} for any given filling fraction.

The quantization of conductance into $e^2/h$ times an integer may be made apparent by considering a discrete pumping cycle rather than a continuous process such as constant applied field leading to a continuous current.\cite{Thouless} In the one-dimensional chain, such a discrete cycle consists of smoothly changing the phase $\phi$ of the CDW by $2\pi$, while in an IQHE cylinder it corresponds to smoothly increasing the flux along the cylinder axis by a single flux quantum.\cite{Laughlin} In both cases, a precisely quantized number of charges, equal to the integer topological quantum number, is transferred between opposing edges of the system after a single pumping cycle.

Although it is one of the central manifestations of topology, the fact that the quantization of charge transport between the edges of a finite system is determined by an integer number characterising a different system without edges, may well seem counterintuitive. It may become even more so once you realize that unlike the Landau-level picture for the IQHE, the topological invariant for the CDW system may be negative as well as positive. In the $1/3$-filled CDW, for example, a single electron is transported every pumping cycle in the direction of sliding, but a $2/3$-filled CDW instead transfers an electron in the direction \emph{opposite} to the sliding. Several more counterintuitive questions naturally arise, including how the electrons cross between edges of the CDW chain, even though the bulk is strictly insulating and the macroscopic distance between the edges suppresses any direct tunneling; or what the amount of charge localized on each edge is, at any given moment during the pumping cycle. To give a clear and intuitive answer to these questions, we will visualize the topological pumping process in the finite CDW chain in real space and real time.

\section{Topological transport}
Although the matrix Hamiltonian of Eq.~\ref{HmfMatrix} conveniently describes the plane-wave electron states in a chain with periodic boundary conditions, it is less convenient for describing what happens at the edges of a chain with open boundaries. Writing the Hamiltonian again in terms of operators $\cdop{j}$ that create localized electrons along a chain with $N$ sites, it becomes:
\begin{align}
\hat{H}_{\text{MF}} &=\left(\hat{c}_{j=0}^{\dagger},\hat{c}_{1}^{\dagger}  \ldots \hat{c}_{N-1}^{\dagger}\right) h \left(\begin{array}{c} 
\hat{c}_{0}\\ \hat{c}_{1}\\ \vdots\\ \hat{c}_{N-1} \end{array}\right), \notag \\
~&~\notag\\
h &= \left(\begin{array}{cccccc} \epsilon_{0}^{\phantom *} & -t & 0 & \ldots & 0 & -\tilde{t} \\
  -t & \epsilon_{1}^{\phantom *} & -t & 0 & \ldots & 0 \\
0 & -t & \ddots & \ddots &  & \vdots \\
\vdots & 0 & \ddots &  &  & 0 \\
0 & \vdots &  &  &  & -t \\
 -\tilde{t} & 0 & \ldots & 0 & -t & \epsilon_{N-1}^{\phantom *} \end{array}\right).
\label{hMatrix}  
\end{align}
Here, we defined $\epsilon_j=2 V \Delta \cos(Qja+\phi)$, and we use the same approximation $\langle \cdop{j-1}\cop{j-1} + \cdop{j+1}\cop{j+1} \rangle \approx 2 \rho(j)$ that was used in the Supplementary Material to construct the Hamiltonian with periodic boundary conditions.\cite{SupMat} The elements $\tilde{t}$ at the corners of the matrix can be used to model different types of connections between the edges of the chain. The periodic boundary conditions considered before correspond to $\tilde{t}=t$. Having an open chain with nothing attached to the edges, is represented by $\tilde{t}=0$. An intermediate case, where the edges are connected via a weak link, corresponding for example to an additional wire in an experimental implementation, may be modeled by taking $0<\tilde{t}\ll t$.
%
\begin{figure}[t]
\includegraphics[width=\columnwidth]{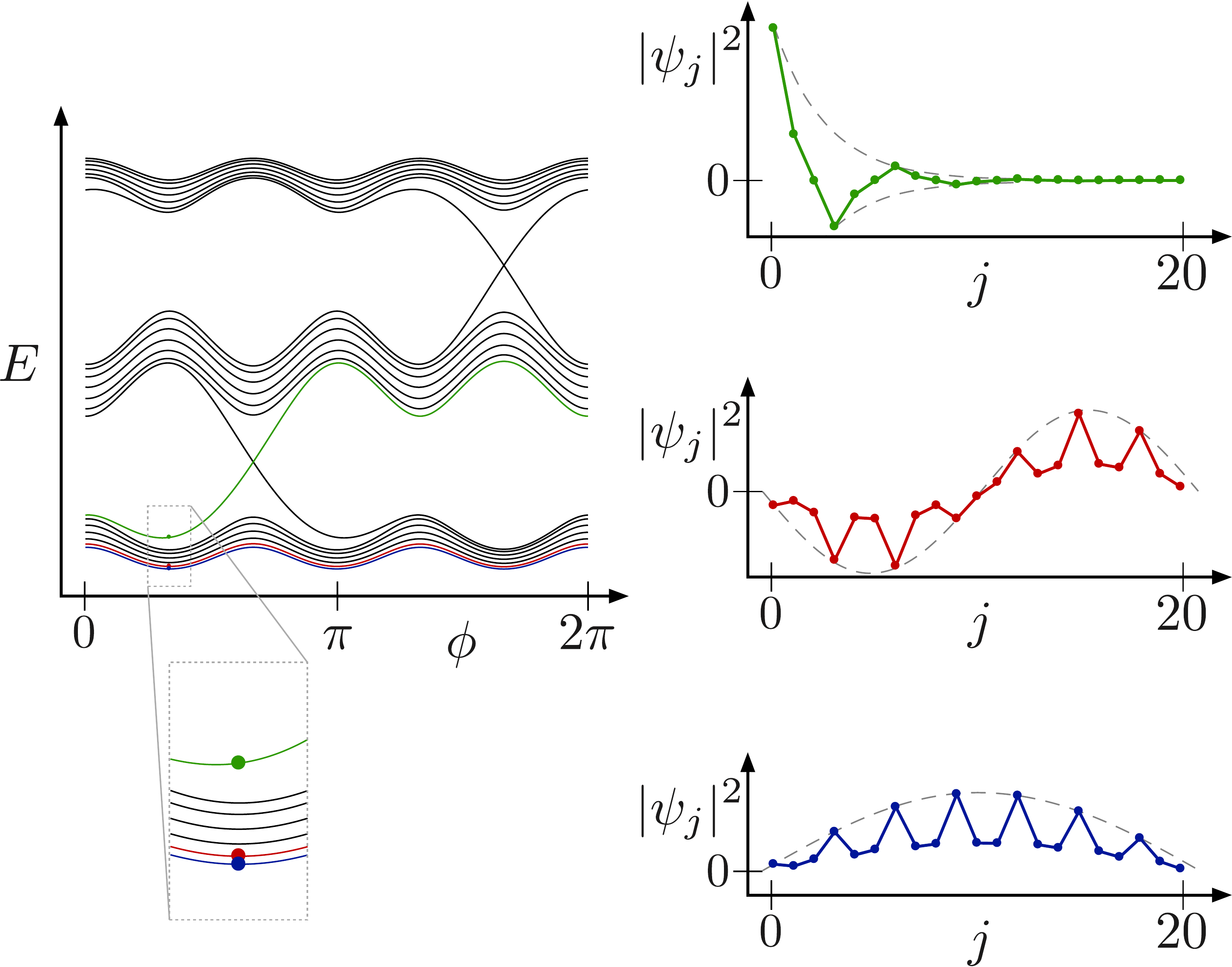}
\caption{\label{fig2}The energies of the electron states in a CDW on an open chain, as a function of the phase variable $\phi$. The wave functions for the lowest energy state, first excited state, and the first edge state are displayed on the right, with colors corresponding to the labels shown in the inset. Also indicated are the exponential and sinusoidal envelopes, which show the low-energy wave functions to be modulated particle-in-a-box states, and the edge state to be exponentially localized at position $j=0$.\\ To create this figure, we used the (arbitrary) parameter values $N=21$, $V \Delta=0.5 t$, and $Q=2\pi/(3a)$. We varied $\phi$ between zero and $2\pi$ in steps of $0.01$.}
\end{figure}
%
\begin{figure}[th]
\includegraphics[width=\columnwidth]{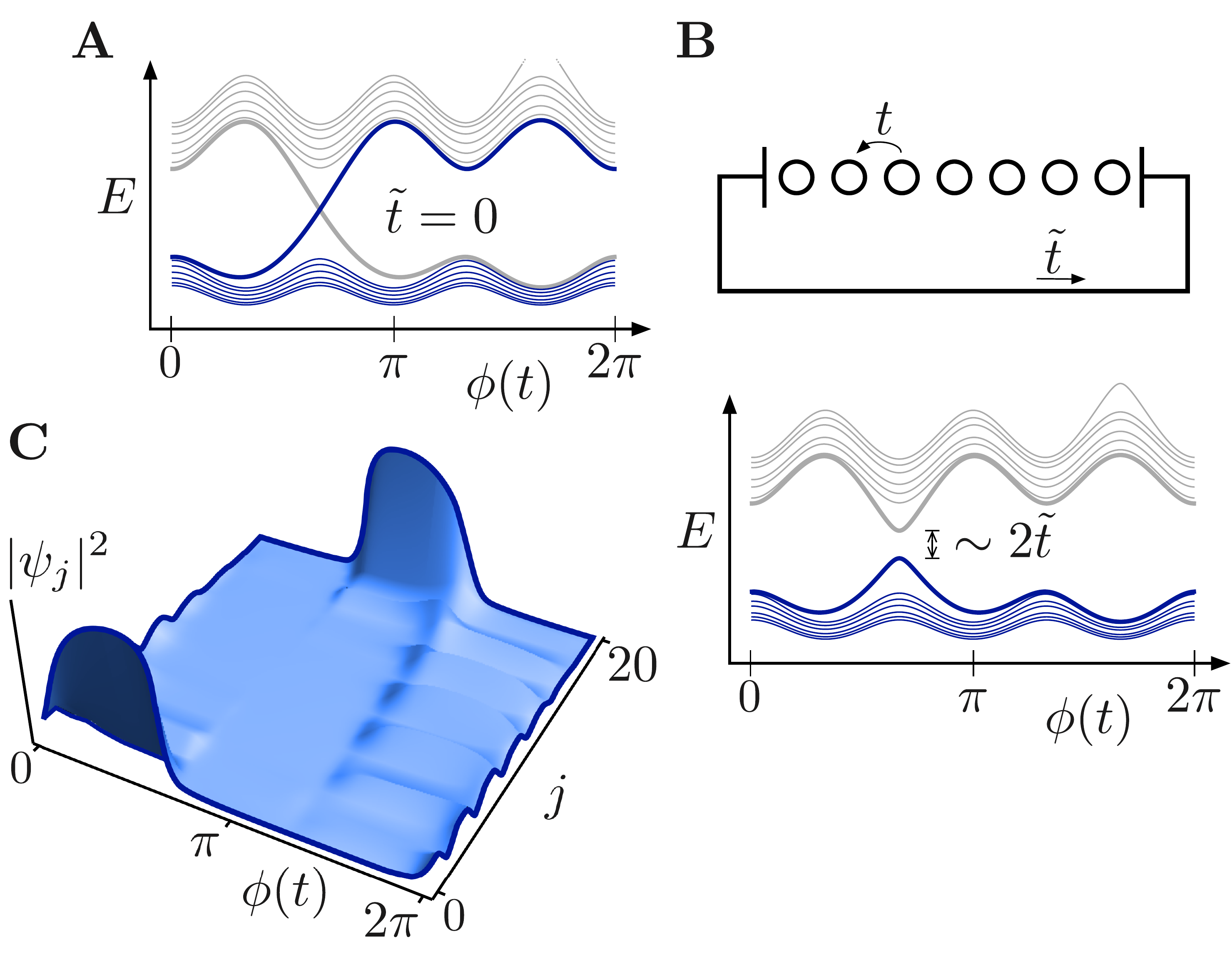}
\caption{\label{fig3}{\bf A} The energies of the electron states in a CDW on an open chain, as a function of the phase variable $\phi(t)$, which may be smoothly varied from zero to $2\pi$ as a function of time. Without any connection between the ends of the chain, the many-particle CDW ground state with one fully occupied band of electronic states at $\phi=0$ will evolve into an excited state in which the lowest band has one empty state and the second band has a single electron in it (indicated in blue). {\bf B} Upon including a weak connection between edges in the model, which corresponds to the wire that would be used to measure the pumped current in any experimental realization of the topological pump, the edge states in the spectrum become gapped (\emph{i.e.} they undergo a so-called avoided crossing). The many-particle ground state now evolves back into itself after a full $2\pi$ cycle of the phase $\phi$ (indicated in blue). While doing so, a single electron traverses the external wire (at $\phi\sim 2\pi/3$) and comes back though the bulk of the CDW (between $\phi \sim 4\pi/3$ and $\phi\sim 2\pi$). {\bf C} The real-space wave function of the lowest energy edge state (thick blue line in panel {\bf B}), as the phase $\phi$ is driven from zero to $2 \pi$.\\
In this figure, we used the (arbitrary) parameter values $N=21$, $V \Delta=0.5 t$, and $Q=2\pi/(3a)$. In panel {\bf B}, we additionally set $\tilde{t}=0.25 t$. The phase $\phi$ was varied in steps of $0.01$.}
\end{figure}

The eigenvalues of the matrix in Eq.~\eqref{hMatrix} may be found numerically for moderate values of $N$. For each value of the phase $\phi$, there are $N$ different eigenvalues, as shown in Fig.~\ref{fig2}. For the system with periodic boundary conditions, these $N$ eigenvalues would correspond to the plane-wave energies found above, labelled by $N$ different values of the wave number $k$. Taking open boundary conditions, they instead correspond to standing wave solutions, with nodes at the edges of the chain, and various numbers of zeroes in between. These can be visualized directly by plotting the eigenvectors of $h$, and as shown in Fig.~\ref{fig2} they indeed look like particle-in-a-box states, modulated by the periodic CDW.

For certain specials values of $\phi$ we also find energy eigenvalues inside what would have been the energy intervals without any solutions for the periodic chain (the so-called band gaps). The corresponding eigenvectors reveal these special states to be exponentially localized at either the right or left edge of the open chain. Two edge states, one localized on the right, and one on the left, cross each other in energy as $\phi$ is varied. The states can be degenerate there, because their exponential localization results in zero wave function overlap between them. Once we add a connection between the edges, taking $\tilde{t}$ to be small but non-zero, the degeneracy will be lifted, as shown in Fig.~\ref{fig3}B.

To visualize the topological transport, we now consider for example the CDW of Fig.~\ref{fig3}A with $N/3$ electrons and $\phi=0$, so that the lowest-energy band of electronic states is fully occupied in the ground state (indicated in blue in Fig.~\ref{fig3}A). All higher-energy bands are empty. Smoothly varying $\phi$ then implies driving it slowly enough for each occupied electron state to remain occupied, and each empty state to stay empty. As $\phi$ grows, the states in the lowest band thus remain occupied. As $\phi$ becomes greater than approximately $\pi/3$, one of the bulk (particle-in-a-box) bands is slowly transformed into a state localized at the left edge of the CDW chain. Since we vary $\phi$ very slowly however, it can remain occupied. In fact, for $\tilde{t}=0$, we can increase $\phi$ all the way to $\phi \gtrsim 2\pi/3$, and find the occupied left edge state going up in energy all the way to the bottom of the second band, while the topmost state of the first band has become unoccupied. After a full cycle, at $\phi=2\pi$, the system is in an excited state, with an electron state occupied in the second band, and a hole present in the first.

Although energy is pumped into the system, this is not yet the quantized electron pump we were hoping for. To usefully employ a device that pumps electrons from one of its edges to another, you need to connect the edges by a wire, and use the flow of electrons through the wire to do work. Such an external connection between the edges can be modelled in the Hamiltonian of Eq.~\eqref{hMatrix} by taking $\tilde{t}$ to be small, but non-zero. A small gap between the energies of the crossing edge states then opens up, as shown in Fig.~\ref{fig3}B. Under the smooth variation of $\phi$, the highest occupied state in the first band then becomes a left edge state at $\phi\approx\pi/3$, then crosses through the external wire as the energies of the two edge states avoid crossing, and comes back at the right edge of the CDW chain for $\phi\approx 2\pi/3$. After a full cycle, the system is back in its initial state, and ready to be used again.

During the cycle, a single electron is pumped from a left CDW edge state to the right, allowing the extraction of work. The entire evolution of the electronic wave function can be followed and visualized as a function of time, as shown in Fig.~\ref{fig3}C. The fact that precisely a single electron is transferred, is due to there being one set of single-electron edge states crossing the first band gap. For a $2/3$-filled CDW chain, three electrons will flow through the wire. Two of them (both in the energy gap between the first and second band) flow in opposite directions, and do not contribute to the overall transport. The single remaining transferred electron (in the energy gap between the second and third band) flows from right to left, and thus constitutes a current in the direction opposite to the sliding motion of the CDW.

\section{Discussion}
The fact that the discrete number of particles pumped during a topological transport cycle in an open chain coincides with a topological quantum number calculated for periodic boundary conditions, is a manifestation of the so-called bulk-boundary correspondence~\cite{Thouless,QAPT}. It can be interpreted by considering the energies shown in Fig.~\ref{fig1}A. Both the particle pumping and the emergence of a nonzero topological quantum number arise from a single phenomenon, namely the so-called inversion of energy-eigenstates. For $\tilde{\Delta}=0$, the energy gaps between the bands would vanish, and the second and third band would touch at one point, $k=0$. The lowest-energy state in the third band with $k$ just below zero then connects smoothly to the highest-energy state in the second band with $k$ just above zero. If we make $\tilde{\Delta}$ non-zero, the connection will change and a gap will open up, in the same way that a gap opened up between the states in Fig.~\ref{fig3}B. The qualitative form of the electronic wave functions just above and just below $k=0$ are not affected by the gap opening at $k=0$, however. The lowest-energy state in the third band therefore connects two qualitatively different states (just above and just below $k=0$) for any non-zero value of $\tilde{\Delta}$. The topological quantum number of the CDW can be thought of as effectively keeping count of the number of such so-called band inversions.\cite{Jorrit2017,jorrit2} In a system with open boundaries, these same inverted states extend into the band gap, and form the exponentially localized states seen in Fig.~\ref{fig3}C.\cite{zak1,zak2,Ana}

If the system is large, so that edge state wave functions do not overlap, their energies will cross inside a band gap. These crossings are exploited during topological transport to transfer a discrete number of charges from one side of the system to the other. Visualizing this process by directly plotting the wave functions of an accessible model system in real space and real time clarifies the nature of the topological transport, and gives an intuitive understanding for why the transport is quantized, why it requires a non-zero topological invariant, and what happens to the electronic wave function of the transported charges throughout the pumping cycle.

The one-dimensional CDW system discussed here has the advantage that it allows a direct visualization of the topological transport. The conclusions, however, are not unique for this system. As we saw before, the mean-field CDW Hamiltonian can be mapped onto a model for the IQHE on a two-dimensional cylinder.\cite{Flicker} The electronic states in the IQHE setup are less straightforward to visualize and follow in time, both because of their two-dimensional nature, and because of the technical requirement of introducing a minimal coupling between momentum and magnetic flux to ensure gauge invariance.\cite{ourPRB} The energies of the electron states in the IQHE, however, are the same as those in Figs.~\ref{fig1} and~\ref{fig2}, with the caveat that $\phi$ now labels canonical momentum and all states in the lowest band, for all values of $\phi$, are simultaneously occupied. Laughlin then showed that under the insertion of an additional flux quantum along the interior of the IQHE cylinder, all states increase their momentum value by one unit,\cite{Laughlin} moving one step in the diagram of Fig.~\ref{fig2}. It is straightforwardly checked that the effect is the same as that of the topological pump discussed here in the CDW context. That is, for $n=1/3$, a single electron moves from being localized along the perimeter at the left end of the cylinder, to the right.

The visualization of topological transport established here for a particularly accessible example system, can be straightforwardly adopted to other models, and gives qualitative insight into the emergence of edge states and quantized adiabatic particle transport in all types of topological insulators.

\subsection*{Appendix: Topological quantum numbers}
\label{appA}
Using for example the Kubo formula, the transverse conductance, $\sigma_{\text{H}}$, of a two-dimensional electronic system with periodic boundary conditions may be written as:\cite{TKNN}
\begin{align}
  \label{Chern}
  \sigma_{\text{H}} &= \frac{e^2}{h} \sum_{m \in \,\text{occ}} c_m  \\
  c_m &= \frac{-i}{2\pi} \int d\vec{k} \int d\vec{x} \, \left( \partial_{k_x} \psi^m_{\vec{k}}(\vec{x}) \right)^* \left( \partial_{k_y} \psi^m_{\vec{k}}(\vec{x})  \right) - c.c. \notag
\end{align}
Here, the spatial integral runs over one unit cell, while $\vec{k}$ cover all inequivalent wave numbers (the first Brillouin zone). The index $m$ labels energy bands, and is summed over occupied bands only. The wave function $\psi^m_{\vec{k}}(\vec{x})$ is the $m^{\text{th}}$ eigenvector of the Hamiltonian matrix $H_k$. The integrand in the final line is known as the Berry curvature for the completely occupied band $m$, and the number $c_m$ is called the Chern number for band $m$. As we will see, the contribution $c_m$ of band $m$ to the conductivity does not depend on the energies of the states in the band. This makes $c_m$ a topological quantum number that is unaffected by any changes of the Hamiltonian, as long as they do not cause bands to cross.

To evaluate $c_m$ for the bands in the model of Eq.~\eqref{HmfMatrix}, we interpret $k$ and $\phi$ to correspond to $k_x$ and $k_y$ coordinates in Eq.~\eqref{Chern}. It may then seem like we need to explicitly find the wave functions $\psi^m_{\vec{k}}(\vec{x})$. In fact, however, the only thing that matters is how the wave function \emph{changes} as $\vec{k}$ is varied, not what it is for any particular momentum. To see the changes in the wave function as a function of $k$ and $\phi$, it suffices to find the wave function at some selected points $(k,\phi)$, and then interpolate smoothly between those.

For concreteness, consider filling fraction $n=1/3$, so that there are three energy bands, with one occupied. First consider the non-interacting model with $\tilde{\Delta}=0$. The matrix $H_k$ is then diagonal, and the three bands correspond to electrons created by $\cdop{k+Q}$, $\cdop{k+2Q}$, and $\cdop{k}$. For non-zero $\tilde{\Delta}$ the matrix has non-zero off-diagonal elements, and the eigenvectors will be superpositions, of the form $A\cdop{k+Q}+B\cdop{k+2Q}+C\cdop{k}$. Focusing first on the lowest, occupied band at the particular point $k=0$, we know that at $\tilde{\Delta}=0$ and for any value of $\phi$, it corresponds to the state with $A=B=0$ and $C=1$. For small, but non-zero $\tilde{\Delta}$, the coefficients $A$ and $B$ become of order $\tilde{\Delta}/E$, where $E$ is the energy separating the lowest band from the higher two bands. Since $E$ is large, we may approximate the lowest band at $k=0$ to correspond purely to $\cdop{k}$, even for small non-zero values of $\tilde{\Delta}$. As will become clear shortly, the small error in assigning this wave function will not affect the value of the Chern number.

Following the same line of reasoning, we can consider the state at $k=-\pi/(3a)$, again for any value of $\phi$. Here, the lowest two bands are degenerate, and will both contribute to the wave function for non-zero $\tilde{\Delta}$. The highest band, however, is again well-separated from the others in energy, and may be ignored to first order in $\tilde{\Delta}/E$. To find the wave function of the lowest energy state, we thus consider only the rows and columns of the matrix $H_k$ associated with $\cdop{k}$ and $\cdop{k+Q}$. This yields a $2$x$2$ matrix whose eigenvectors can be straightforwardly found, giving a lowest energy state with $A=-e^{-i\phi}\sqrt{1/2}$, $B=0$, and $C=\sqrt{1/2}$. Repeating the same arguments at $k=\pi/(3a)$ gives a wave function there with $A=0$, $B=-e^{i\phi}\sqrt{1/2}$, and $C=\sqrt{1/2}$.

It now turns out to be impossible to define smooth functions $A(k,\phi)=\tilde{A}(k)e^{-i\phi}$, $B(k,\phi)=\tilde{B}(k)e^{i\phi}$, and $C(k,\phi)=\tilde{C}(k)$ that reduce to the obtained values for $A$, $B$, and $C$ at the three $k$-points spanning the Brillouin zone. What is possible, however, is to divide the Brillouin zone into two regions, $k\in [-\pi/(3a),0]$ and $k\in [0,\pi/(3a)]$, and find two sets of functions that interpolate smoothly between the end points for each region individually. The topological nature of the Chern number then becomes immediately apparent when we write Eq.~\eqref{Chern} in terms of the interpolating functions:
\begin{align}
  c_1 = \frac{-i}{2\pi} \int_0^{2\pi} d\phi &\left[ -2i \int_{-\pi/(3a)}^0 dk \frac{\partial \tilde{A}(k)}{\partial k} \tilde{A}(k) \right. \notag \\
  & ~~~\left.+ 2i \int_0^{\pi/(3a)} dk\, \frac{\partial \tilde{B}(k)}{\partial k} \tilde{B}(k) \right] \notag \\
  = \frac{-1}{\pi} \int_0^{2\pi} d\phi & \left[ \int_{\tilde{A}(-\pi/(3a))}^{\tilde{A}(0)} \tilde{A} d\tilde{A} - \int_{\tilde{B}(0)}^{\tilde{B}(\pi/(3a))} \tilde{B} d\tilde{B} \right] \notag \\
  = 1. ~~\,\hphantom{ \int_0^{2\pi} d\phi}&
\end{align}
Notice that we could evaluate the integrals without ever specifying the precise form of the functions $\tilde{A}(k)$ and $\tilde{B}(k)$. This shows what it means for the number $c_1$ to be topological: as long as the functions $\tilde{A}(k)$ and $\tilde{B}(k)$ have the correct values at the $k$-points where a gap is created, the Chern number is completely insensitive to how we interpolate between these end points. We can thus freely change the Hamiltonian and its eigenvectors (described by $\tilde{A}(k)$ and $\tilde{B}(k)$), as long as the changes do not cause any additional gaps to open or close.

A similar calculation will show that the Chern number $c_2$ for the second band equals $-2$, while that for the uppermost band is again $1$ (and the sum of Chern numbers for all bands equals zero, as it should since a completely filled band structure cannot transport any charge). In terms of quantized transport between edges in an open CDW chain, this means that for filling fraction $n=1/3$ sliding the CDW over a single wave length results in a single electron being transferred in the direction of sliding. For a filling of $n=2/3$ we have $c_1+c_2=-1$, so that a single electron will be transferred in the direction \emph{opposite} to the sliding motion. 

~

\subsection*{Acknowledgments}
This work is part of the Delta Institute for Theoretical Physics (DITP) consortium, a program of the Netherlands Organization for Scientific Research (NWO) that is funded by the Dutch Ministry of Education, Culture and Science (OCW).\\
~\\
~\\
~\\
~\\

\newpage
~
\newpage

\onecolumngrid
\section*{Supplemental Material}
\noindent
In this supplementary material, we present the detailed steps involved in the calculations of the main text. We hope these may serve as a convenient starting point for formulating exercises, questions, and simulations connected to the visualization of topological transport.

~

\twocolumngrid
\subsection{The mean-field Hamiltonian}\label{sec1}
To find the mean-field description of the charge-density wave (CDW) in a one-dimensional chain, we start from the Hamiltonian:
\begin{align}
  \hat{H} = \sum_{j=0}^{N-1} \left\{ -t \left(\cdop{j} \cop{j+1} + \cdop{j+1} \cop{j} \right)+ V \cdop{j} \cop{j} \cdop{j+1} \cop{j+1} \right\}.
  \label{Hcdw}
\end{align}
Recall that the operators $\cdop{j}$ and $\cop{j}$ respectively create and annihilate an electron (ignoring spin) at position $x = j a$, where $a$ is the lattice constant and $j$ an integer site label. Notice that for now, we will use periodic boundary conditions, so that the labels $j$ and $j+N$ correspond to the same site. The parameter $t>0$ signifies the likelihood for an electron to tunnel between neighboring sites, and $V$ is the strength of the nearest-neighbor Coulomb interaction.

The idea behind the mean-field analysis is that we expect to be able to make a reasonable guess for the ground state expectation value of the electron density $\rho(j) \equiv \langle \cdop{j} \cop{j} \rangle$. At the end of the calculation, one may check that the initial guess is indeed consistent with the model we end up with. To see how we can use the fact that we know what to expect for the density operator, first rewrite it as:
\begin{align}
  \cdop{j} \cop{j} &= \langle \cdop{j} \cop{j} \rangle + \hat{f}_j \notag \\
  \text{with}~~  \hat{f}_j &\equiv \cdop{j} \cop{j} - \langle \cdop{j} \cop{j} \rangle.
\label{fluctop}
\end{align}
This expression defines the fluctuation operator $\hat{f}_j$, and does not involve any approximation yet. Assuming that we do have a good guess for the expectation value of the electron density however, we may assume that the expectation value of the fluctuations $\hat{f}_j$ is small, and its square even smaller. We can use this by rewriting the Hamiltonian in terms of the fluctuation operator, and then neglecting all terms of quadratic (or higher) order in the fluctuations:
\begin{align}
  \hat{H} &= \sum_{j=0}^{N-1} \left\{ -t (\cdop{j} \cop{j+1} +\cdop{j+1}\cop{j}) \right. \notag\\
  &~~~~~ \left. +V \left(\langle \cdop{j} \cop{j} \rangle + \hat{f}_j\right) \left(\langle \cdop{j+1} \cop{j+1} \rangle + \hat{f}_{j+1}\right) \right\} \notag \\
          &\approx \sum_{j=0}^{N-1} \left\{ -t  (\cdop{j} \cop{j+1} +\cdop{j+1}\cop{j}) \right. \notag\\
          &~~~~~ \left. +V \left(\langle \cdop{j} \cop{j} \rangle \hat{f}_{j+1} + \langle \cdop{j+1} \cop{j+1} \rangle \hat{f}_j \right. \right. \notag \\
  &~~~~~~~~\left.\left. + \langle \cdop{j} \cop{j} \rangle \langle \cdop{j+1} \cop{j+1} \rangle  \right) \right\}
\end{align}
In the final line, we can use Eq.~\eqref{fluctop} to write the remaining fluctuation operators in terms of density operators again, and then replace the expectation values $\langle \cdop{j} \cop{j} \rangle$ with the guessed electron density $\rho(j)$.
\begin{align}
  \hat{H} &\approx \sum_{j=0}^{N-1} \left\{ -t (\cdop{j} \cop{j+1} +\cdop{j+1}\cop{j}) \right. \notag\\
  &~~~~~ \left. +V \left(\rho(j) \cdop{j+1} \cop{j+1} + \rho(j+1) \cdop{j} \cop{j} \right. \right. \notag \\
  &~~~~~~~~\left.\left. - \rho(j)\rho(j+1) \right)  \right\} \notag \\
  &= \sum_{j=0}^{N-1} \left\{ -t (\cdop{j} \cop{j+1} + t \cdop{j+1} \cop{j})  \right. \notag\\
  &~~~~~ +V \left(\rho(j-1) + \rho(j+1)\right) \cdop{j} \cop{j}  \notag \\
  &~~~~~~~~\left. -  2 V \rho(j)\rho(j+1)  \right\}. 
\end{align}
In the second line, we used the periodic boundary conditions to shift the summation index in one of the terms. The final term in this Hamiltonian, $-2V\rho(j)\rho(j+1)$ is a constant that can be removed by a suitable redefinition of the zero of energy (which is always arbitrary). Since the charge density is a smooth and continuous function, we can also use the approximation $\langle \cdop{j-1}\cop{j-1} + \cdop{j+1}\cop{j+1} \rangle \approx 2 \rho(j)$ to find the final form of the mean-field CDW Hamiltonian:
\begin{align}
  \hat{H} &\approx \sum_{j=0}^{N-1} \left\{ -t (\cdop{j} \cop{j+1} +\cdop{j+1} \cop{j} )+ 2V \rho(j) \cdop{j} \cop{j} \right\} \notag \\
  & =  \hat{H}_{\text{MF}}.
  \label{HcdwMF}
\end{align}
In this mean-field Hamiltonian, we can now explicitly write our guess for the electron density: $\rho(j) = \rho_0 + \Delta \cos(Q j a + \phi)$, with $\Delta$ the CDW amplitude, $Q$ the CDW wave number, and $\phi$ its phase. The terms in the final mean-field Hamiltonian proportional to the average electron density $\rho_0$ are all of the form $\sum_j V \rho_0 \cdop{j}\cop{j}$. They thus act as a chemical potential, and may be ignored from here on, because the value of the chemical potential can be chosen at the end of the calculation to produce the correct average number of electrons. Finding and minimizing the ground state energy, in principle allows us to check that the guessed form of $\rho(j)$ is self-consistent. Since it is not essential for the CDW topology, we will not elaborate on the self-consistency conditions here.

\subsection{Alternative boundary conditions}
Notice that the mean-field Hamiltonian of Eq.~\eqref{HcdwMF} can immediately be put into matrix form:
\begin{align}
\hat{H}_{\text{MF}} &=\left(\hat{c}_{j=0}^{\dagger},\hat{c}_{1}^{\dagger}  \ldots \hat{c}_{N-1}^{\dagger}\right) h \left(\begin{array}{c} 
\hat{c}_{0}\\ \hat{c}_{1}\\ \vdots\\ \hat{c}_{N-1} \end{array}\right), \notag \\
~&~\notag\\
h &= \left(\begin{array}{cccccc} \epsilon_{0}^{\phantom *} & -t & 0 & \ldots & 0 & -\tilde{t} \\
  -t & \epsilon_{1}^{\phantom *} & -t & 0 & \ldots & 0 \\
0 & -t & \ddots & \ddots &  & \vdots \\
\vdots & 0 & \ddots &  &  & 0 \\
0 & \vdots &  &  &  & -t \\
 -\tilde{t} & 0 & \ldots & 0 & -t & \epsilon_{N-1}^{\phantom *} \end{array}\right).
\label{hMatrix}  
\end{align}
Here, we defined $\epsilon_j=2 V \rho(j) = 2 V \Delta \cos(Qja+\phi)$. The elements $\tilde{t}$ at the corners of the matrix are equal to $t$ for the periodic boundary conditions studied so far. They can also be used to implement different boundary conditions. Having an open chain with nothing attached to the edges, for example, is described by using $\tilde{t}=0$. An intermediate case, where the edges are connected via a weak link, corresponding for example to an additional wire in an experimental implementation, may be modeled by taking $0<\tilde{t}\ll t$.

\subsection{The Fourier-transformed Hamiltonian}
Assuming periodic boundary conditions again, we can introduce creation and annihilation operators for electrons whose wave functions are plane waves:
\begin{align}
  \cdop{k} &= \sqrt{1/N} \sum_{j=0}^{N-1} e^{-ikja} \cdop{j}, \notag \\
  \cop{k} &= \sqrt{1/N}  \sum_{j=0}^{N-1} e^{ikja} \cop{j}.
\end{align}
Here, the variable $k$ signifies the wave number of the plane wave. Notice that the plane-wave electrons can be interpreted as a (discrete) Fourier transformation of the localized electrons. Because of the assumed periodicity of the chain in real space, the plane wave amplitude at position $j$ has to be equal to that on position $j+N$. This implies that the wave number $k$ is only allowed to have discrete values $2\pi m/(Na)$, with $m$ an integer. Moreover,  because the chain consists of discrete atoms, and the electronic wave function can only have a non-zero value at atomic positions, the wave numbers turn out to be periodic. This means that $\cdop{k=2\pi/a}$ creates the same electronic wave function (that is, it has the same amplitude on each of the discrete atomic sites) as $\cdop{k=0}$. For the one-dimensional chain then, $k$ can be allowed to have discrete values $2\pi m/(Na)$, with $m\in\{0,1,2,\dots,N-1\}$.

We can also express the localized electrons in terms of the plane-wave ones:
\begin{align}
  \cdop{j} &= \sqrt{1/N} \sum_{0 \le k < 2 \pi / a} e^{ikja} \cdop{k}, \notag \\
  \cop{j} &= \sqrt{1/N}  \sum_{0 \le k < 2 \pi / a} e^{-ikja} \cop{k}.
\end{align}
These definitions can be substituted directly into the Hamiltonian of Eq.~\eqref{HcdwMF}:
\begin{align}
  \hat{H}_{\text{MF}} &= \frac{1}{N} \sum_j \sum_k \sum_{k'}  \left[ -t e^{ikja} e^{-ik'(j+1)a} \right.\notag \\
                        &~~~~~ \left. - te^{ik(j+1)a} e^{-ik'ja}  \right.\notag \\
                        &~~~~~ \left.+ 2V \Delta \cos(Q j a + \phi) e^{ikja} e^{-ik'ja}\right] \cdop{k} \cop{k'}.
\end{align}
In these expressions, we can write the cosine as a sum of exponentials:
\begin{align}
  2 \cos(Qja+\phi) = e^{i(Qja+\phi)} + e^{-i(Qja+\phi)}.
\end{align}
Substituting the definition of the delta function $\delta_{k,k'} = 1/N \sum_j e^{i(k-k')ja}$, then allows us to perform the sum over one of the momenta:
\begin{align}
  \hat{H}_{\text{MF}} &= \frac{1}{N} \sum_k \sum_{k'}  \left( -t \delta_{k,k'} e^{-ik'a} - t \delta_{k,k'} e^{ika}  \right.\notag \\
                        &~~~~~ \left.+ V \Delta e^{i \phi}\delta_{k+Q,k'} +V \Delta e^{-i\phi} \delta_{k-Q,k'}\right) \cdop{k} \cop{k'} \notag \\
  &= \sum_{0 \le k < 2 \pi / a} \left\{ -2t \cos(ka) \cdop{k} \cop{k}   \right.\notag \\
                        &~~~~~ \left.+ V \Delta e^{i \phi} \cdop{k} \cop{k+Q}  +V \Delta e^{-i\phi} \cdop{k} \cop{k-Q} \right\}.
\end{align}
This is the same form for the Fourier transformed Hamiltonian as that used in the main text.

\subsection{Writing the Hamiltonian in matrix form}\label{sec4}
To facilitate the use of numerical software for calculating the eigenvalues of the Hamiltonian, it is convenient to write it in matrix form:
\begin{align}
\hat{H}_{\text{MF}} &=\sum_{k} \left(\hat{c}_{k+Q}^{\dagger},\hat{c}_{k+2Q}^{\dagger}  \ldots \hat{c}_{k+qQ}^{\dagger}\right)H_{k}\left(\begin{array}{c} 
                                                                                                                                          \hat{c}_{k+Q}\\ \hat{c}_{k+2Q}\\ \vdots\\ \hat{c}_{k+qQ} \end{array}\right). \notag
\end{align}
Here, we assumed $Q=n \cdot 2\pi/a$, with $n=p/q$ a co-prime fraction, so that periodic boundary conditions imply $\cdop{k+qQ}=\cdop{k}$. Writing the Hamiltonian this way, however, one should be careful with the sum over wave numbers. If we simply sum $k$ over the values $2\pi m/(Na)$, with $m\in\{0,1,2,\dots,N-1\}$, electrons with momentum equal to for example $3Q/2$ will be created both by the first component $\cdop{k+Q}$ of the vector of creation operators (for $k=Q/2$), and by the final component $\cdop{k+qQ}$ in the vector (for $k=3Q/2$). The eigenvalues of the Hamiltonian matrix $H_k$, however, will correspond directly to the energies of $\hat{H}_{\text{MF}}$ only if it is expressed in an orthonormal basis, or equivalently, if the every possible electron state is created only once within the sum over $k$. This can be achieved by restricting the range of momentum values summed over to $2\pi m/(Na)$, with $m\in\{0,1,2,\dots,N/q-1\}$.\footnote{Notice that the restricted range of momenta consists precisely of all momenta within the so-called reduced Brillouin zone that corresponds to the enlarged real-space unit cell of size $qa$ in the CDW state.}

To find the matrix $H_k$ for the one-dimensional chain, we thus first rewrite the equation for $\hat{H}_{\text{MF}}$ such that it contains only a sum over this restricted range of momentum values:
\begin{align}\label{MFH}
\hat{H}_{\text{MF}} &= \sum_{0 \le k < 2 \pi / a} \left\{ \epsilon_k \cdop{k} \cop{k}   \right.\notag \\
                    &~~~~~~~~ \left.+ V \Delta e^{i \phi} \cdop{k} \cop{k+Q}  +V \Delta e^{-i\phi} \cdop{k} \cop{k-Q} \right\} \notag \\
                    &=\sum_{0\le k < 2\pi /qa} ~\sum_{n=1}^q \left\{ \epsilon_{k+nQ} \cdop{k+nQ} \cop{k+nQ} \right. \notag \\
                    &~~~~~~~~ + V \Delta  e^{i \phi}  \cdop{k+nQ} \cop{k+(n+1)Q} \notag \\
                    &~~~~~~~~ \left.+ V \Delta  e^{-i \phi}  \cdop{k+(n+1)Q} \cop{k+nQ} \right\}. 
\end{align}
Notice that in this expression, we used $(q+1)Q=Q$, which follows from the periodicity of the wave numbers.

Writing out this equation in the desired matrix form, yields the final expression for  $H_k$:
\begin{align}
H_{k} &= \left(\begin{array}{cccccc} \epsilon_{k+Q}^{\phantom *} & \tilde{\Delta}^{\phantom *} & 0 & \ldots & 0 & \tilde{\Delta}^{*} \\
 \tilde{\Delta}^{*} & \epsilon_{k+2Q}^{\phantom *} & \tilde{\Delta}^{\phantom *} & 0 & \ldots & 0 \\
0 & \tilde{\Delta}^{*} & \ddots & \ddots &  & \vdots \\
\vdots & 0 & \ddots &  &  & 0 \\
0 & \vdots &  &  &  & \tilde{\Delta}^{\phantom *} \\
 \tilde{\Delta}^{\phantom *} & 0 & \ldots & 0 & \tilde{\Delta}^{*} & \epsilon_{k+qQ}^{\phantom *} \end{array}\right).
\label{HmfMatrix}  
\end{align}
Recall that here, $\tilde{\Delta}=V\Delta e^{i \phi}$, and $k+qQ=k$ owing to the periodicity of the wave numbers. Note that this form of the matrix does not apply to the special case $q=2$, for which the mean field Hamiltonian of Eq.~\eqref{MFH} reduces to just $\hat{H}_{\text{MF}}=\sum_k \epsilon_k \cdop{k}\cop{k} + V \Delta \cos(\phi) \cdop{k}\cop{k+\pi}$. At the value $\phi=\pi/2$, the band gap in this $q=2$ Hamiltonian closes, and adiabatic transport is no longer possible. We therefore do not consider this special case.

\subsection{Suggested exercises}
\begin{itemize}
\item Reproduce the derivations in sections~{A}-{D} above.
\item Make an animation of the topological transport in the mean-field CDW chain. That is, plot the wave function amplitude $|\psi_j|^2$ of the occupied state with the highest energy in Fig. 3 of the main text, for some initial value of the phase $\phi$. Then animate cyclic variations of $\phi$. Do this for different values of the model parameters (in particular $\tilde{t}/t$).
\item (\emph{more advanced}) Include weak impurities in the model by adding a potential energy $E_{\text{imp}}=\sum_j \eta(j) \cdop{j}\cop{j}$ to the Hamiltonian, where for every $j$, $\eta(j)$ is an independent random number between $-t/5$ and $t/5$. Observe the effect of the impurities on eigenfunctions in the bulk (they typically become more localized), while the topological transport is unaffected (quantized transport still occurs).
\end{itemize}

~\\

~\\

~\\

~\\

\end{document}